\documentclass[aps,prl,preprintnumbers,amsmath,amssymb,floatfix,nofootinbib,twocolumn]{revtex4-1}

\usepackage{dcolumn}
\usepackage{bm}
\usepackage[dvips]{graphicx}
\usepackage{epsfig}
\usepackage{amsmath, amsfonts, amssymb}

\newcommand{\ie}{{\em i.e. }}

\newcommand{\be}{\begin{equation}}
\newcommand{\ee}{\end{equation}}
\newcommand{\bee}{\begin{equation*}}
\newcommand{\eee}{\end{equation*}}
\newcommand{\bea}{\begin{eqnarray}}
\newcommand{\eea}{\end{eqnarray}}
\newcommand{\bean}{\begin{eqnarray*}}
\newcommand{\eean}{\end{eqnarray*}}

\usepackage{color}

\begin{document}

\title{Echoes of the Electroweak Phase Transition:\\ Discovering a second Higgs doublet through $A_0 \rightarrow H_0 Z$.}

\author{G. C. Dorsch, S. Huber, K. Mimasu and J. M. No}

\affiliation{{\it Department of Physics and Astronomy, University of Sussex, 
BN1 9QH Brighton, United Kingdom}}

\date{\today}

\begin{abstract}
The existence of a second Higgs doublet in Nature could lead to a cosmological first order electroweak phase transition and explain the origin of the matter-antimatter asymmetry in the Universe. We obtain the spectrum and properties of the new scalars $H_0$, $A_0$ and $H^{\pm}$ that signal such a phase transition, and show that the observation of the decay $A_0 \rightarrow H_0 Z$ at LHC would be a `smoking gun' signature of these scenarios. We analyze the LHC search prospects for this decay in the $\ell \ell b\bar{b}$ and $\ell \ell W^{+} W^{-}$ final states, arguing that current data may be sensitive to this signature in the former channel as well as there being great potential for a discovery in either one at the very early stages of the 14 TeV run.
\end{abstract}

\maketitle

A key goal of the Large Hadron Collider (LHC) physics programme is to reveal the structure of the sector responsible for electroweak symmetry breaking (EWSB) in Nature. While ongoing analyses show that the properties of the newly discovered Higgs particle~\cite{Aad:2012tfa, Chatrchyan:2012ufa} are close to those expected for the Standard Model (SM) Higgs boson $h$, it still needs to be determined whether the scalar sector consists of one $SU(2)_L$ doublet or has a richer structure, containing additional states. Scenarios with extra scalar doublets are very well-motivated, naturally arising both in the context of weakly coupled completions of the SM, like the Minimal Supersymmetric Standard Model (MSSM) (for a review of EWSB in the MSSM, see \cite{Djouadi:2005gj}) and its extensions, and in strongly coupled ones, such as Composite Higgs scenarios \cite{Mrazek:2011iu}. Moreover, simple extensions of the SM scalar sector like Two-Higgs-Doublet-Models (2HDMs) could address an important open question at the 
interface of particle physics and cosmology, namely the generation of the cosmic matter-antimatter asymmetry, via Electroweak Baryogenesis \cite{Turok:1990zg,Cline:1995dg,Fromme:2006cm,Cline:2011mm}.

The implications of 2HDMs for electroweak physics and LHC searches have been widely studied in the literature (see, {\it e.g.} \cite{Branco:2011iw,Celis:2013rcs,Krawczyk:2013gia,Grinstein:2013npa,Chen:2013rba,Craig:2013hca, Eberhardt:2013uba,Kanemura:2009mk,Baglio:2014nea,Coleppa:2014hxa}). 
An important aspect, with a big impact on searches for the second Higgs doublet at LHC, is the mass spectrum of the new states $S_i$: 
a charged scalar $H^{\pm}$ and two neutral scalars $H_0$, $A_0$. In the MSSM, and generically in 2HDMs arising from weakly coupled completions of the SM, 
the mass splittings $\Delta m_{i}$ among these scalars are small, $\Delta m_{i} \ll v$, with $v = 246$ GeV the electroweak (EW) scale. The heavier the new scalar states, the more compressed their mass spectrum. In particular, for the MSSM the splittings $\Delta m_{i}$ are smaller than the mass of the gauge bosons $W^{\pm}$ and $Z$, and the decays $S_i \rightarrow Z S_j$ or $S_i \rightarrow W^{\pm} S_j$ are not kinematically allowed. The observation of these decays at LHC would point to a very different kind of 2HDM. However, while these decays have already been considered \cite{Kanemura:2009mk,Coleppa:2014hxa}, such scenarios remain largely unexplored and are not currently a part of the main LHC search programme.

In this letter we show that the decay $A_0 \rightarrow Z H_0$ is the signature of a strongly first order electroweak phase transition (EWPT) in 2HDMs, as needed for Electroweak Baryogenesis. We then show that, while current 2HDM by ATLAS and CMS collaborations 
are not tailored to probing these scenarios, searches for the decay $A_0 \rightarrow Z H_0$ are promising at the 14 TeV run of LHC in the $\ell \ell\,b\bar{b}$ and $\ell \ell\,W^{+} W^{-}$ channels. This signature could then provide a connection between the generation of the cosmic matter-antimatter asymmetry in the Early Universe and searches for new physics at LHC. 

The letter is organized as follows: in section I we briefly review 2HDMs and analyze how the strength of the EWPT depends on parameter space for these models. 
In section II we discuss the connection between the decay $A_0 \rightarrow Z H_0$ and the strength of the EWPT, and in section III we analyze the LHC search prospects 
in the $\ell \ell\,b\bar{b}$ and $\ell \ell\,W^{+} W^{-}$ channels. We conclude in section IV.

\vspace{-5mm}

\subsection{I. 2HDMs and the Electroweak Phase Transition}

\vspace{-3mm}

The scalar sector of a 2HDM contains two scalar doublets $\Phi_{1,2}$, and its physical spectrum consists of a charged Higgs $H^{\pm}$, two Charge-Parity (CP)-even scalars $h$ and $H_0$, and a CP-odd scalar $A_0$ (we assume, for simplicity, no CP violation in the scalar sector). We identify the lightest CP-even Higgs $h$ with the scalar resonance recently observed at the LHC, {\it i.e.} we fix $m_h = 125$~GeV. The remaining physical parameters in the scalar potential are: the masses $m_{H_0}$, $m_{A_0}$, $m_{H^{\pm}}$; two angles $\beta$ and $\alpha$, the former begin related to the ratio of vacuum expectation values (vev) of the two scalar doublets $v_{1,2}$ (with $v^2_1 + v^2_2 = v^2$) and the latter related to mixing between CP-even states; and a dimensionful parameter $\mu$ (for a review of 2HDMs, see {\it e.g.} \cite{Branco:2011iw}). We define $\alpha$ such that, when $\alpha = \beta$, the state $H_0$ decouples from gauge bosons, and $h$ 
has SM-like properties (the \emph{alignment limit}; see \cite{Dorsch:2013wja} for a more detailed discussion). 

In general the existence of two scalar doublets $\Phi_{1,2}$ coupling to fermions opens an undesirable window for flavour changing neutral currents at tree-level. This can be avoided by imposing a $Z_2$ symmetry, softly broken by the $\mu$ parameter in the scalar potential, which forces each fermion type to couple to one doublet only \cite{Glashow:1976nt}. By convention, up-type quarks always couple to the second doublet, but which doublet couples to leptons and down-type quarks may vary. We will focus here on the so-called Type I model, in which all fermions couple to the same doublet. Another scenario is the so-called Type II model, where down-type quarks and leptons couple to a different doublet from up-type quarks, and of which the scalar sector of the MSSM is a particular instance.
 
In order to study the strength of the EWPT in 2HDMs, we perform a Monte Carlo scan over a wide range of $m_{H_0}$, $m_{A_0}$, $m_{H^{\pm}}$, $\tan\beta$, 
$\alpha - \beta$ and $\mu$ using an in-house numerical code developed in~\cite{Dorsch:2013wja}. The code is interfaced to 2HDMC~\cite{Eriksson:2009ws} and HiggsBounds~\cite{Bechtle:2013wla} to select points in parameter space that satisfy unitarity, perturbativity, electroweak precision constraints and collider bounds. Stability of the potential is checked at 1-loop level by requiring that the electroweak minimum (\ie the one for which $v_1^2+v_2^2=v^2$) be the deepest minimum of the effective potential~\cite{Dorsch:2013wja}. As for flavor constraints, for the Type I model the only relevant one\footnote{The points excluded by other constraints, in particular $B^0-\bar{B^0}$ mixing and $Z\rightarrow b\bar{b}$, are also excluded by $b\rightarrow s\gamma$.}~comes from $b\rightarrow s\gamma$, which we take into account~\cite{Mahmoudi:2009zx}. 
In addition, the measured properties of $h$, impose further constraints on $\tan\beta$ and $\alpha-\beta$ (see {\it e.g.} \cite{Chen:2013rba}). We note that, while the type of 2HDM considered is irrelevant for the EWPT (since all types couple in the same way to the top quark), it does affect constraints from colliders and Higgs properties. We choose a Type I 2HDM, which is less constrained than a Type II. The results below show that a strong EWPT prefers a SM-like state $h$, and thus 2HDMs with a strong EWPT also satisfy Type II constraints. 

A point in our scan satisfying the above constraints is called a \emph{physical point}. For each of them, the strength of the EWPT is computed via the thermal 1-loop effective potential by increasing the temperature until a point is reached when the potential has two degenerate minima, which then defines the critical temperature $T_c$. The phase transition is considered 
strong when $v_c/T_c \ge 1$ \cite{Moore:1998swa,Quiros:1994dr}, with $v_c$ the magnitude of the broken vev at $T_c$ (see \cite{Dorsch:2013wja} for details). 

\begin{figure}[ht]
\begin{center}
	\includegraphics[width=0.48\textwidth, clip]{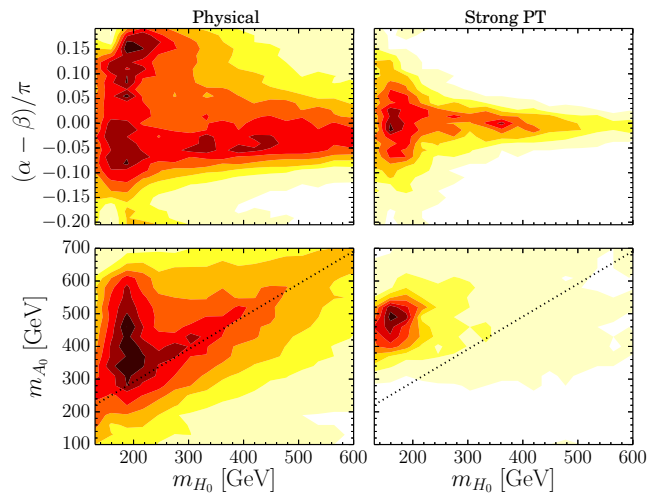}
	\vspace{-7mm}
	\caption{\small  Heat-maps for the physical region (\emph{left}) and region with a strongly first order EWPT (\emph{right}). \emph{Top}: ($m_{H_0}, \alpha - \beta$)-plane. \emph{Bottom}: ($m_{H_0}, m_{A_0}$)-plane. The dotted-black line corresponds to $m_{A_0} = m_{H_0} + m_{Z}$.}
	\vspace{-7mm}
	\label{fig:EWPT}
	\end{center}
\end{figure}

In Figure~\ref{fig:EWPT} we show the heat-maps of physical points (\emph{left}) and points with a strongly first order EWPT (\emph{right}) in the planes ($m_{H_0}, \alpha - \beta$) and ($m_{H_0}, m_{A_0}$).
Altogether, a strong EWPT, as needed for Electroweak Baryogenesis, favours the light Higgs state $h$ to have SM-like properties, {\it i.e.} small $\alpha-\beta$ and moderate $\tan\beta$ \cite{Dorsch:2013wja,DHMN}. The range of $\alpha-\beta$ leading to a strong EWPT shrinks as the CP-even state $H_0$ becomes heavier. This can be understood from the fact that away from the alignment limit $\alpha \simeq \beta$, both $h$ and $H_0$ ``share" the vev $v$, {\it i.e.} both participate in the EWPT, and the phase transition becomes weaker as the states participating in it get heavier (see {\it e.g.} \cite{Quiros:1994dr}). In addition, Figure~\ref{fig:EWPT} shows that a strong 
EWPT in 2HDMs scenarios strongly favours a rather heavy CP-odd scalar state $A_0$ ($m_{A_0} > 300$ GeV), together with a large mass splitting $m_{A_0} - m_{H_0} \gtrsim v$.

As we discuss in the next section, these results point towards the $A_0 \rightarrow Z H_0$ decay channel as a `smoking gun' signature of 2HDMs with a strong EWPT, to be searched for at the upcoming 14 TeV run of the LHC.

\vspace{-5mm}

\subsection{II. The Decay $A_0 \rightarrow Z\,H_0$}

\vspace{-3mm}

Current 2HDM searches at LHC are mainly motivated by the MSSM, where scalar mass splittings are dictated by the gauge couplings and do not exceed $m_Z$. The decays $S_i \rightarrow Z S_j$ (for $S_i \in {H_0, A_0}$) are not kinematically allowed, and ATLAS/CMS searches are thus not tailored to them. Most searches so far have focused on $H_0\rightarrow W^+W^-$ \cite{ATLAS:2013zla,TheATLAScollaboration:2013zha} and $H_0\rightarrow Z Z$ \cite{ATLAS:2013nma,CMS:2013pea}, or on the search of the CP-odd state via $A_0 \rightarrow \tau^+ \tau^-$~\cite{ATLAS:2012doa} and $A_0 \rightarrow Z h$~\cite{CMS:2013eua,ATLAS:2013pub}.

Our results from the previous section show, however, that a strong EWPT in 2HDMs strongly favours a heavy CP-odd state $A_0$ with a mass splitting $m_{A_0}-m_{H_0} \gtrsim v$, in addition to $\alpha \sim \beta$ (although a small departure from this limit is allowed). The decay $A_0 \rightarrow Z H_0$ is then strongly favoured both due to the large amount of phase space available, and because the coupling $g_{A_0ZH_0} \sim \mathrm{cos}(\alpha-\beta)$ is unsuppressed in the alignment limit. In contrast, the coupling $g_{A_0Zh} \sim \mathrm{sin}(\alpha-\beta)$ vanishes in that limit, and the decay $A_0 \rightarrow Z h$ is further suppressed due to $A_0 \rightarrow Z H_0$ remaining dominant away from alignment (see Figure 2). 

The competing decay channels would then be $A_0 \rightarrow t\bar{t}$ and possibly $A_0 \rightarrow W^{\pm} H^{\mp}$. The former is suppressed as $(\tan\beta)^{-2}$, which is moderate in the scenarios under discussion, and is anyway subdominant for $m_{A_0}-m_{H_0} \gtrsim v$ (Figure 2, \emph{top}). The latter will depend on the splitting $m_{A_0}-m_{H^{\pm}}$. Electroweak precision observables require the charged Higgs $H^{\pm}$ to be relatively close in mass to either $H_0$ or $A_0$ \cite{Grimus:2007if}, and as a result the decay $A_0 \rightarrow W^{\pm} H^{\mp}$ will be either kinematically forbidden or similar in magnitude to $A_0 \rightarrow Z H_0$. Our scan for Type I scenario does not show a strong preference for one case over the other. For Type II, we expect that $A_0 \rightarrow W^{\pm} H^{\mp}$ will be generically disfavoured compared to $A_0 \rightarrow Z H_0$, since the mass of $H^{\pm}$ is constrained to be $m_{H^{\pm}} > 360$ GeV at 95\% C.L. \cite{Hermann:2012fc}. For the rest of the 
analysis, we assume for simplicity $m_{H^{\pm}} \sim m_{A_0}$ (for $m_{H^{\pm}} \sim m_{H_0}$, the decay $A_0 \rightarrow W^{\pm} H^{\mp}$ may suppress the decay branching fraction for $A_0 \rightarrow Z H_0$ by a factor $\sim 2$ \cite{DHMN}).

\begin{figure}[ht]
\begin{center}
	\includegraphics[width=0.38\textwidth, clip]{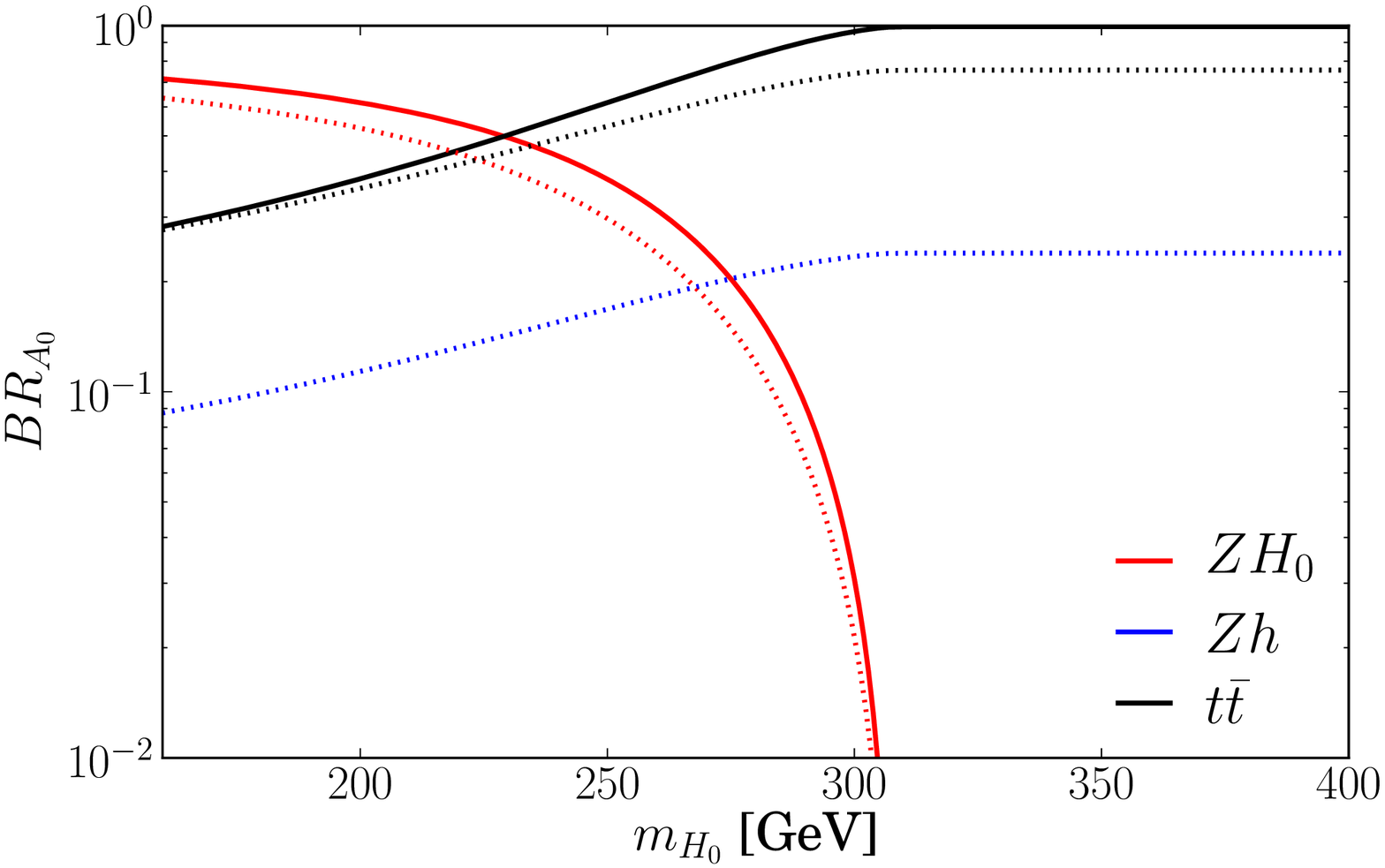}\\
	\includegraphics[width=0.38\textwidth, clip]{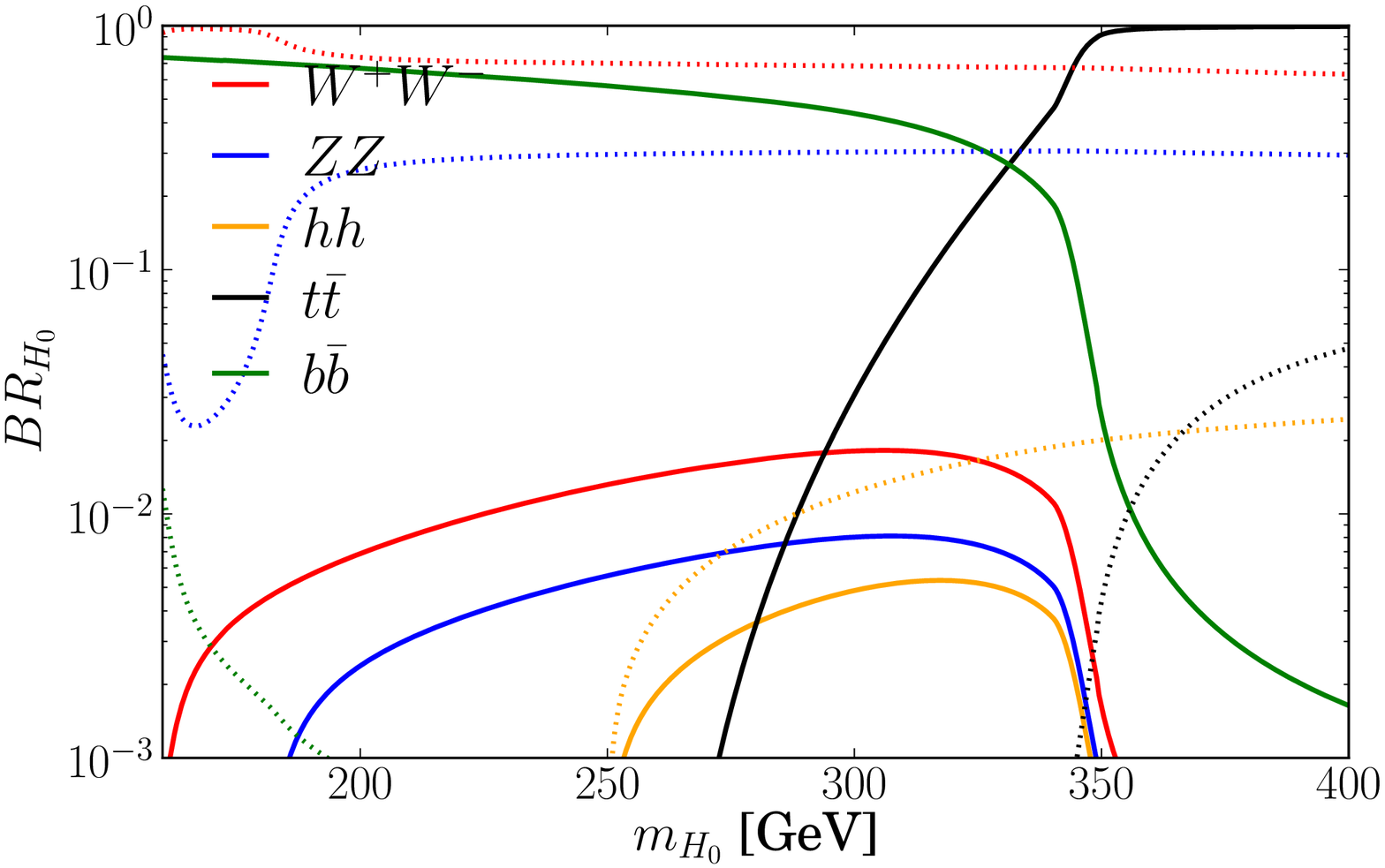}
    \vspace{-2mm}
	\caption{\small \emph{Top}: Main Branching Ratios of the CP-odd scalar $A_0$ as a function of $m_{H_0}$ for $m_{A_0} = m_{H^{\pm}} = 400$~ GeV, $\tan\beta = 2$, $\mu = 100$~GeV, $\alpha - \beta = 0.001\pi$ (solid lines) and $\alpha - \beta = 0.1\pi$ (dotted lines). \emph{Bottom}:  Main Branching Ratios of $H_0$ as a function of $m_{H_0}$ (same benchmark parameters as in \emph{top}).}
	\vspace{-5mm}
	\label{fig:Br}
	\end{center}
\end{figure}

In Figure~\ref{fig:Br} (\emph{top}) we show the main decay branching fractions of $A_0$ as a function of $m_{H_0}$ for two benchmark points, which represent two prototypical scenarios, henceforth referred to as \emph{A} and \emph{B}: $m_{A_0} = m_{H^{\pm}} = 400$~GeV, $\mu = 100$~GeV, $\mathrm{tan}\beta = 2$, with $(\alpha - \beta) = 0.001\,\pi$ and $(\alpha - \beta) = 0.1\,\pi$ respectively. We observe that for $m_{A_0} - m_{H_0} \gtrsim v$, $A_0 \rightarrow Z H_0$ largely dominates over $A_0 \rightarrow t\bar{t}$ and $A_0 \rightarrow Z h$ in both cases. This behaviour is even more pronounced as we approach the alignment limit. 

As for the subsequent decay of $H_0$, in the alignment limit the decays $H_0\rightarrow W^+W^-$, $H_0\rightarrow Z Z$ and $H_0 \rightarrow h h$ are suppressed, which for the case of a relatively light $H_0$, as preferred by a strong EWPT (Figure 1, \emph{right}), leaves $H_0 \rightarrow b \bar{b}$ as the dominant decay mode, as shown in Figure 2 (\emph{bottom}). As we move away from $\alpha = \beta$, the decays into gauge bosons and $h$ become more important, and for $m_{H_0} < 250$ GeV, the decay $H_0\rightarrow W^+W^-$ quickly dominates. For $m_{H_0} \gtrsim 250$ GeV and small $\mu$ this is still the case, while increasing $\mu$ enhances the coupling $g_{H_0hh}$ \cite{Kanemura:2004mg} and consequently $H_0 \rightarrow h h$, when kinematically allowed.

The above discussion highlights the fact that for 2HDMs with a strongly first order EWPT, the main search channel at LHC will either be $p p \rightarrow A_0 \rightarrow Z H_0 \rightarrow \ell \ell b \bar{b}$ or $p p \rightarrow A_0 \rightarrow Z H_0 \rightarrow \ell \ell W^+W^-$, depending on how close the 2HDM is to the alignment limit. These two alternative scenarios are exemplified by our two benchmark choices, described above. 

\begin{figure}[ht]
\begin{center}
	\includegraphics[width=0.48\textwidth, clip]{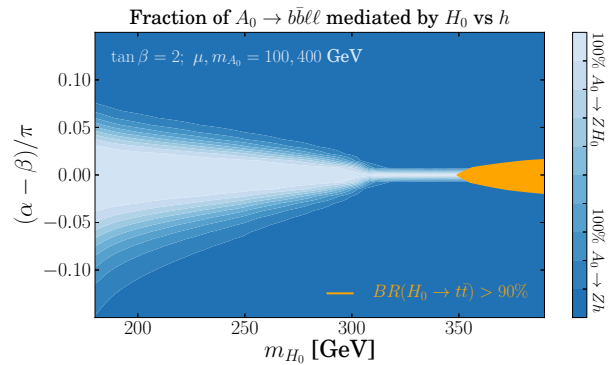}
	\vspace{-6mm}
	\caption{\small Fraction of $\ell \ell b \bar{b}$ events from $Z h$ vs $Z H_0$ decays, for $m_{A_0} = m_{H^{\pm}} = 400$~GeV, $\mu = 100$~GeV, $\mathrm{tan}\beta = 2$.}
	\vspace{-5mm}
	\label{fig:Br2}
	\end{center}
\end{figure}

Before we turn to the search prospects in the $\ell \ell b \bar{b}$ and $\ell \ell W^+W^-$ channels at the 14 TeV run of the LHC for these two benchmark scenarios, let us remark that these prototypical scenarios generically evade direct $H_0$ searches. On one side, for $\alpha \sim \beta$, the decay 
$H_0 \rightarrow b \bar{b}$ is difficult to extract from the QCD background. On the other, as we deviate from the alignment limit with $\alpha > \beta$, and for moderate 
$\mathrm{tan}\beta$, the $H_0$ coupling to top quarks decreases and direct production of $H_0$ in gluon fusion gets very suppressed, evading searches in $W^+W^-$ and $Z Z$
channels. 

\vspace{-5mm}

\subsection{III. LHC Search for $A_0$ in $\ell\ell\,b\bar{b}$ and $\ell\ell\,W^{+}W^{-}$}

\vspace{-3mm}

We now determine the search prospects for the scenario discussed above at the LHC using simple `cut and count' analyses. We first consider the $\ell \ell b \bar{b}$ final state (see also \cite{Coleppa:2014hxa}). 
This channel is favoured for the nearly aligned benchmark scenario, $A$, for which we choose $m_{H_0} = 180$~GeV.
As shown in Figure 3, the decay $A_0 \rightarrow Z H_0 \rightarrow \ell \ell b \bar{b}$ largely dominates over $A_0 \rightarrow Z h \rightarrow \ell \ell b \bar{b}$, which becomes extremely suppressed and may be safely disregarded. 

The analysis described here was applied to both the 8 and 14 TeV energy stages of the LHC. It was found that the 8 TeV run could potentially observe the process, our analysis yielding $S/\sqrt{S+B}\sim 3$ for 20 fb$^{-1}$ of integrated luminosity. We interpret this as the fact that even the lower energy LHC data could begin to explore regions of 2HDM  parameter space conducive to a strongly first order EW phase transition. For example, the analysis in~\cite{TheATLAScollaboration:2013lia} could be reinterpreted or even modified to search for this decay mode with intermediate $m_{A_0}$. In this letter, we choose to present the 14 TeV analysis, which permits a clear discovery at the very early stages of the forthcoming run.

The main SM backgrounds to $\ell \ell b \bar{b}$ are: (i) QCD $t\bar{t}$ production (with $t\bar{t} \rightarrow b W^{+} \bar{b} W^{-} \rightarrow b \ell^{+} \nu_{\ell} \bar{b} \ell^{-} \bar{\nu}_{\ell}$), (ii) $Z b\bar{b}$ production (with $Z \rightarrow \ell \ell$), (iii) $Z Z$ production (with $Z \rightarrow \ell \ell, \, Z \rightarrow b\bar{b}$) and (iv) production of the light Higgs $h$ in association with a $Z$ boson (with $h \rightarrow b\bar{b}$ and $Z \rightarrow \ell \ell$). We implement the Type-I 2HDM in {\sc FeynRules} \cite{Christensen:2008py,Degrande:2011ua}, and use {\sc MadGraph5$\_$aMC$@$NLO} \cite{Alwall:2011uj,Alwall:2014hca} to generate our signal and background analysis samples. These samples are passed on to {\sc Pythia} \cite{Sjostrand:2007gs} for parton showering and hadronization, and then to {\sc Delphes} \cite{deFavereau:2013fsa} for a detector simulation. We rescale the cross sections for our signal and two dominant backgrounds, $Zb\bar{b}$ and $t\bar{t}$, estimating their 
respective NLO values via $K$-factors: $K\simeq 1.6$ for the 
signal (which we compute using {\sc Sushi} \cite{Harlander:2012pb}), $K\simeq 1.4$ for $Z b\bar{b}$ \cite{Campbell:2005zv} and $K\simeq 1.5$ for $t\bar{t}$, noting that the remaining backgrounds become negligible upon signal selection.

\begin{table}[ht]
\begin{tabular}{c| c | c | c | c | c |}

& Signal & $t\bar t$ & $Z\, b\bar b$ & $Z Z$  & $Z\, h$ \\
\hline 
&  &  &  &  &\\ [-2ex]
Event selection& 14.6 & 1578 & 424 & 7.3 & 2.7 \\ [0.5ex]
$80$ $< m_{\ell\ell} < 100$ GeV& 13.1 & 240 & 388 & 6.6  &2.5 \\ [0.5ex]
$\begin{array}{c}
H_T^\mathrm{bb} > 150 \,\mathrm{GeV} \\
H_T^\mathrm{\ell\ell bb} > 280\, \mathrm{GeV}
 \end{array}$
& 8.2 & 57 & 83 &  0.8 &0.74 \\ [2ex]
$\Delta R_{bb} < 2.5$, $\Delta R_{\ell\ell} < 1.6$& 5.3 & 5.4 & 28.3 & 0.75  &0.68 \\ [0.5ex]
\hline &  &  &  &  &\\ [-2ex]
$m_{bb}$, $m_{\ell\ell bb}$ signal region& 3.2 & 1.37 & 3.2 & $< 0.01$ &$< 0.02$ \\ [0.5ex]
\hline
\end{tabular}
\caption{\small Event selection (see section III) and background reduction in the $\ell \ell b \bar{b}$ final state. We show the NLO cross section (in fb) after successive cuts for the signal 
$A_0 \rightarrow Z H_0 \rightarrow \ell \ell b \bar{b}$ and the dominant backgrounds $t \bar{t}$ and $Z b \bar{b}$, while $Z Z$ and $Z h\rightarrow \ell \ell b \bar{b}$ are shown at LO.}\vspace{-4mm}

\end{table}

\begin{figure}[ht]
\begin{center}
	\includegraphics[width=0.48\textwidth, clip]{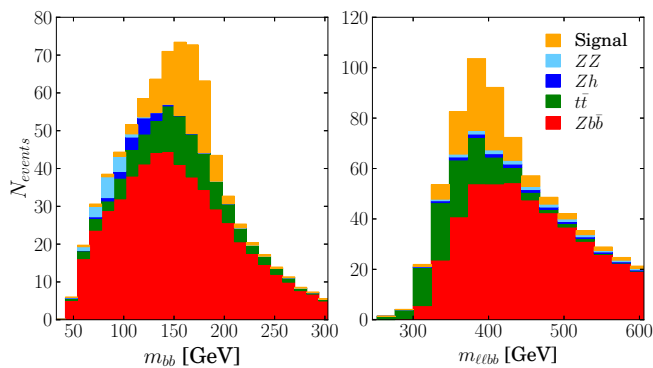}
	\vspace{-5mm}
	\caption{\small $m_{bb}$ (\emph{left}) and $m_{\ell\ell bb}$ (\emph{right}) distributions after analysis cuts, with the various contributions stacked (for an integrated luminosity of $\mathcal{L} = 20\,\, \mathrm{fb}^{-1}$).}
	\vspace{-5mm}
	\label{fig:dist1}
	\end{center}
\end{figure}

For event selection, we require the presence of two isolated (within a cone of 0.3) same flavour (SF) leptons in the final state with $P^{\ell_1}_{T} > 40$, $P^{\ell_2}_{T} > 20$ and $\left| \eta_{\ell} \right| < $ 2.5 (2.7) for electrons (muons), together with two b-tagged\footnote[2]{We estimate the kinematical dependence of the b-tagging efficiency from 
a recent CMS performance note \cite{CMS:2013vea}. Our acceptance region is divided into ($\eta,P_T$) categories
according to Figure 15 of \cite{CMS:2013vea}. The efficiency imposed on a jet belonging to a given category is obtained as the average of the $\eta$ and $P_T$ bins, whose values range between 60 - 70\%.} jets in the event with $P^{b_1}_{T} > 40$, $P^{b_2}_{T} > 20$ and $\left| \eta_{b} \right| < $ 2.5. Our analysis is presented in Table I, where the rescaled cross sections (in fb) for the signal (S) and SM backgrounds (B) are shown after successive cuts. The $m_{bb}$ and $m_{\ell\ell bb}$ distributions after cuts are shown in Figure 4, with the various contributions stacked (for an integrated luminosity of $\mathcal{L} = 20\, \mathrm{fb}^{-1}$). We define the signal region as $m_{bb} = (m_{H_0} - 20) \pm 30$ GeV and $m_{\ell\ell bb} = (m_{A_0} - 20) \pm 40$ GeV (a small energy loss from the b-jets due to showering is expected). Given the results from Table I, we expect that a discovery value $S/\sqrt{S+B} = 5$ can be obtained with $\mathcal{L}\sim 20$ fb$^{-1}$, considering only statistical uncertainties, 
while assuming a conservative 20\% uncertainty in the background predictions yields the same significance with 40 fb$^{-1}$.

\begin{figure}[ht]
\begin{center}
	\includegraphics[width=0.45\textwidth, clip]{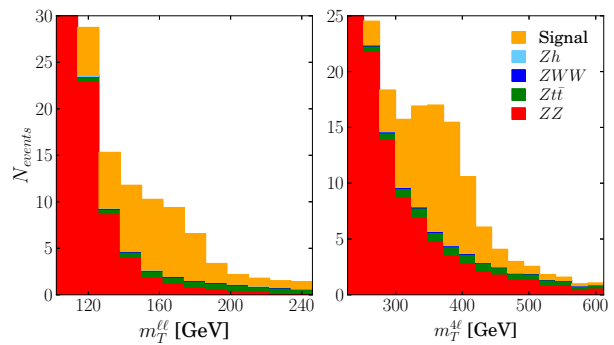}
	\vspace{-2mm}
	\caption{\small $m^{\ell\ell}_{T}$ (\emph{left}) and $m^{4\ell}_{T}$ (\emph{right}) distributions after event selection, with the various contributions stacked (for an integrated luminosity of $\mathcal{L} = 40\, \,\mathrm{fb}^{-1}$).}
	\vspace{-5.5mm}
	\label{fig:dist2}
	\end{center}
\end{figure}

Away from the alignment limit, signal $\ell \ell b \bar{b}$ final states come mostly from $A_0 \rightarrow Z h$ (Figure \ref{fig:Br2}), since Br($H_0 \rightarrow b \bar{b}$) $\ll 1$. Still, for $m_{H_0} \lesssim 250$ GeV the decay $A_0 \rightarrow Z h$ is suppressed, which altogether renders the search in the  $\ell \ell b \bar{b}$ final state challenging. In this regime, $A_0 \rightarrow Z H_0$ with $H_0 \rightarrow W W \rightarrow \ell \nu_{\ell} \ell \nu_{\ell}$ and $Z \rightarrow \ell \ell$ provides the best discovery prospects, as we analyze below for our benchmark $B$ ($H_0 \rightarrow Z Z \rightarrow \ell \ell q \bar{q}$ is also promising \cite{Coleppa:2014hxa}, although the required luminosity for a discovery is much larger).

The main irreducible SM background to $\ell \ell W W$ with two leptonic $W$-decays is diboson ($Z Z$) production with $Z Z \rightarrow \ell \ell \ell \ell$.
Other backgrounds, such as $Z t \bar{t}$, $Z W W$ and $Z h$ yield a combined cross section which is $ < \sigma_{\mathrm{Signal}}/4$ after event selection.
We follow the same selection and analysis procedure as for the $\ell \ell b \bar{b}$ final state, except for requiring the presence of four isolated leptons 
(in two SF pairs) in the final state with $P^{\ell_1}_{T} > 40$ GeV, $P^{\ell_2,\ell_3,\ell_4}_{T} > 20$ GeV. We further require at least one of the SF lepton pairs to reconstruct $m_Z$ within $20$ GeV. The (LO) cross sections at LHC 14 TeV after event selection for the signal, $Z Z$ background and the combined rare backgrounds ($Z t \bar{t}$, $Z W W$ and $Z h$) are respectively $0.93$ fb, $5.6$ fb and $0.25$ fb (a $\Delta R_{\ell\ell}$ cut or a $Z$-veto on the remaining SF lepton pair would further suppress the $Z Z$ background). Again, we rescale the signal and dominant background by their respective NLO $K$-factors (1.35 for $Zt\bar{t}$~\cite{Lazopoulos:2008de} and 1.2 for $ZZ$~\cite{Ohnemus:1994ff}). Defining the transverse mass variables $m^{\ell\ell}_{T}$ and $m^{4\ell}_{T}$
\begin{eqnarray}
\left(m^{\ell\ell}_{T}\right)^2 =  \left(\sqrt{p^2_{T,\ell\ell}+m^2_{\ell\ell}}  + \slash\hspace{-2mm} p_{T}\right)^2 - \left(\vec{p}_{T,\ell\ell} + \slash\hspace{-2mm}\vec{p}_{T}\right)^2 \nonumber \\
m^{4\ell}_{T} = \sqrt{p^2_{T,\ell'\ell'}+m^2_{\ell'\ell'}} + \sqrt{p^2_{T,\ell\ell}+\left(m^{\ell\ell}_{T}\right)^2}
\nonumber
\end{eqnarray}
($\ell'\ell'$ are the two SF leptons most closely reconstructing $m_Z$), a signal region of $m^{4\ell}_{T} > 260$ GeV (see Figure \ref{fig:dist2}) allows to extract a clean signal without any further selection as well as reconstruct the masses of the two particles. A final signal cross section of 1.41 fb compared to a background of 1.7 fb reaches a significance of 5 with $\mathcal{L}\sim 40$ fb$^{-1}$, which increases to 60 fb$^{-1}$ when assuming a 20\% systematic uncertainty on the background prediction.

\vspace{-5mm}

\subsection{IV. Discussion and Outlook}

\vspace{-3mm}

Uncovering the structure of the SM scalar sector will be a central task for the LHC in the coming years. This will have important implications for our understanding of the mechanism of electroweak symmetry-breaking and possibly for open cosmological problems such as the origin of visible matter. Extensions of the SM scalar sector that address these questions may yield distinctive signatures at the LHC. We have shown in this letter that the decay $A_0 \rightarrow Z H_0$ is a `smoking gun' signature of 2HDM scenarios with a strongly first order electroweak phase transition, that could explain the origin of the matter-antimatter asymmetry in the Universe. We claim that current 8 TeV LHC data may begin to be sensitive to such scenarios in the $\ell \ell b \bar{b}$ channel. Furthermore we demonstrate that search prospects for these scenarios at the 14 TeV run in both the previous channel as well as $\ell \ell W W \rightarrow \ell \ell \ell \ell$ are very promising, with a discovery of the new scalar states $A_0$ 
and $H_0$ possible with $\mathcal{L}\sim 20-40$ and $40-60$ fb$^{-1}$ respectively, thus providing a probe of electroweak cosmology at the LHC. 

\vspace{-2mm}

\begin{center}
\textbf{Acknowledgements} 
\end{center}

\vspace{-2mm}

We thank V. Sanz, M. Ramsey-Musolf and S. Su for useful discussions and comments.
S.H., K.M. and J.M.N. are supported by the Science Technology and Facilities
Council (STFC) under grant ST/J000477/1. 
G.C.D. is supported by CAPES (Brazil) under grant 0963/13-5.



\begin{thebibliography}{99}

\bibitem{Aad:2012tfa}
 G.~Aad {\it et al.} [ATLAS Collaboration],
 Phys.\ Lett.\ B {\bf 716} (2012) 1
 [arXiv:1207.7214 [hep-ex]].

\bibitem{Chatrchyan:2012ufa}
 S.~Chatrchyan {\it et al.} [CMS Collaboration],
 Phys.\ Lett.\ B {\bf 716} (2012) 30
 [arXiv:1207.7235 [hep-ex]].

\bibitem{Djouadi:2005gj} 
  A.~Djouadi,
  Phys.\ Rept.\  {\bf 459}, 1 (2008)
  [hep-ph/0503173]. 
 
\bibitem{Mrazek:2011iu}
  J.~Mrazek, A.~Pomarol, R.~Rattazzi, M.~Redi, J.~Serra and A.~Wulzer,
  Nucl.\ Phys.\ B {\bf 853} (2011) 1
  [arXiv:1105.5403 [hep-ph]].

\bibitem{Turok:1990zg} 
  N.~Turok and J.~Zadrozny,
  Nucl.\ Phys.\ B {\bf 358}, 471 (1991).  

\bibitem{Cline:1995dg} 
  J.~M.~Cline, K.~Kainulainen and A.~P.~Vischer,
  Phys.\ Rev.\ D {\bf 54}, 2451 (1996)
  [hep-ph/9506284].  
  
\bibitem{Fromme:2006cm}
  L.~Fromme, S.~J.~Huber and M.~Seniuch,
  JHEP {\bf 0611} (2006) 038
  [hep-ph/0605242].  

\bibitem{Cline:2011mm} 
  J.~M.~Cline, K.~Kainulainen and M.~Trott,
  JHEP {\bf 1111}, 089 (2011)
  [arXiv:1107.3559 [hep-ph]].    
  
  
\bibitem{Branco:2011iw} 
  G.~C.~Branco, P.~M.~Ferreira, L.~Lavoura, M.~N.~Rebelo, M.~Sher and J.~P.~Silva,
  Phys.\ Rept.\  {\bf 516}, 1 (2012)
  [arXiv:1106.0034 [hep-ph]].  

\bibitem{Celis:2013rcs} 
  A.~Celis, V.~Ilisie and A.~Pich,
  JHEP {\bf 1307}, 053 (2013)
  [arXiv:1302.4022 [hep-ph]].


\bibitem{Krawczyk:2013gia} 
  M.~Krawczyk, D.~Sokolowska and B.~Swiezewska,
  J.\ Phys.\ Conf.\ Ser.\  {\bf 447}, 012050 (2013)
  [arXiv:1303.7102 [hep-ph]].

\bibitem{Grinstein:2013npa} 
  B.~Grinstein and P.~Uttayarat,
  JHEP {\bf 1306}, 094 (2013)
  [Erratum-ibid.\  {\bf 1309}, 110 (2013)]
  [arXiv:1304.0028 [hep-ph]].
   
\bibitem{Mahmoudi:2009zx}
  F.~Mahmoudi and O.~Stal,
  Phys.\ Rev.\ D {\bf 81} (2010) 035016
  [arXiv:0907.1791 [hep-ph]].

\bibitem{Chen:2013rba} 
  C.~-Y.~Chen, S.~Dawson and M.~Sher,
  Phys.\ Rev.\ D {\bf 88}, 015018 (2013)
  [arXiv:1305.1624 [hep-ph]].  

\bibitem{Craig:2013hca}
  N.~Craig, J.~Galloway and S.~Thomas,
  arXiv:1305.2424 [hep-ph].

\bibitem{Eberhardt:2013uba} 
  O.~Eberhardt, U.~Nierste and M.~Wiebusch,
  JHEP {\bf 1307}, 118 (2013)
  [arXiv:1305.1649 [hep-ph]].  
  
\bibitem{Kanemura:2009mk}
  S.~Kanemura, S.~Moretti, Y.~Mukai, R.~Santos and K.~Yagyu,
  Phys.\ Rev.\ D {\bf 79} (2009) 055017
  [arXiv:0901.0204 [hep-ph]].
  
  \bibitem{Baglio:2014nea} 
  J.~Baglio, O.~Eberhardt, U.~Nierste and M.~Wiebusch,
  arXiv:1403.1264 [hep-ph].  
  
\bibitem{Coleppa:2014hxa} 
  B.~Coleppa, F.~Kling and S.~Su,
  arXiv:1404.1922 [hep-ph].    
  

  
\bibitem{Dorsch:2013wja}
  G.~C.~Dorsch, S.~J.~Huber and J.~M.~No,
  JHEP {\bf 1310} (2013) 029
  [arXiv:1305.6610 [hep-ph]].

\bibitem{Glashow:1976nt} 
  S.~L.~Glashow and S.~Weinberg,
  Phys.\ Rev.\ D {\bf 15}, 1958 (1977).  
  
\bibitem{Eriksson:2009ws} 
  D.~Eriksson, J.~Rathsman and O.~Stal,
  Comput.\ Phys.\ Commun.\  {\bf 181}, 189 (2010)
  [arXiv:0902.0851 [hep-ph]].  
  
\bibitem{Bechtle:2013wla}
  P.~Bechtle, O.~Brein, S.~Heinemeyer, O.~Stål, T.~Stefaniak, G.~Weiglein and K.~E.~Williams,
  Eur.\ Phys.\ J.\ C {\bf 74} (2014) 2693
  [arXiv:1311.0055 [hep-ph]].

  \bibitem{DHMN} 
  G. C. Dorsch, S. Huber, K. Mimasu and J. M. No,
  Work in Progress.
  \bibitem{Moore:1998swa}
    G.~D.~Moore,
    Phys.\ Rev.\ D {\bf 59} (1999) 014503
    [hep-ph/9805264].
  \bibitem{Quiros:1994dr} 
  M.~Quiros,
  Helv.\ Phys.\ Acta {\bf 67}, 451 (1994).

\bibitem{ATLAS:2013zla} 
  [ATLAS Collaboration],
  ATLAS-CONF-2013-027.
  
\bibitem{TheATLAScollaboration:2013zha} 
  [ATLAS Collaboration],
  ATLAS-CONF-2013-067.  
  
 \bibitem{ATLAS:2013nma} 
  [ATLAS Collaboration],
  ATLAS-CONF-2013-013.
  
 \bibitem{CMS:2013pea} 
  [CMS Collaboration],
  CMS-PAS-HIG-12-024. 

\bibitem{ATLAS:2012doa} 
  [ATLAS Collaboration],
  ATLAS-CONF-2012-094.  
  
\bibitem{CMS:2013eua}
  [CMS Collaboration],
  CMS-PAS-HIG-13-025.    
  
\bibitem{ATLAS:2013pub} 
[ATLAS Collaboration],
ATL-PHYS-PUB-2013-016  
  
\bibitem{Grimus:2007if} 
  W.~Grimus, L.~Lavoura, O.~M.~Ogreid and P.~Osland,
  J.\ Phys.\ G {\bf 35}, 075001 (2008)
  [arXiv:0711.4022 [hep-ph]]; 
  Nucl.\ Phys.\ B {\bf 801}, 81 (2008)
  [arXiv:0802.4353 [hep-ph]].  
 
\bibitem{Hermann:2012fc} 
  T.~Hermann, M.~Misiak and M.~Steinhauser,
  JHEP {\bf 1211}, 036 (2012)
  [arXiv:1208.2788 [hep-ph]]. 
 
\bibitem{Kanemura:2004mg} 
  S.~Kanemura, Y.~Okada, E.~Senaha and C.~-P.~Yuan,
  Phys.\ Rev.\ D {\bf 70}, 115002 (2004)
  [hep-ph/0408364]. 
  
\bibitem{TheATLAScollaboration:2013lia}
  The ATLAS collaboration,
  ATLAS-CONF-2013-079.
  
\bibitem{Christensen:2008py} 
  N.~D.~Christensen and C.~Duhr,
  Comput.\ Phys.\ Commun.\  {\bf 180}, 1614 (2009)
  [arXiv:0806.4194 [hep-ph]].
  
  
\bibitem{Degrande:2011ua} 
  C.~Degrande, C.~Duhr, B.~Fuks, D.~Grellscheid, O.~Mattelaer and T.~Reiter,
  Comput.\ Phys.\ Commun.\  {\bf 183}, 1201 (2012)
  [arXiv:1108.2040 [hep-ph]].  
  
\bibitem{Alwall:2011uj} 
  J.~Alwall, M.~Herquet, F.~Maltoni, O.~Mattelaer and T.~Stelzer,
  JHEP {\bf 1106}, 128 (2011)
  [arXiv:1106.0522 [hep-ph]].  

\bibitem{Alwall:2014hca} 
  J.~Alwall, R.~Frederix, S.~Frixione, V.~Hirschi, F.~Maltoni, O.~Mattelaer, H.~-S.~Shao and T.~Stelzer {\it et al.},
  arXiv:1405.0301 [hep-ph].  
  
\bibitem{Sjostrand:2007gs} 
  T.~Sjostrand, S.~Mrenna and P.~Z.~Skands,
  Comput.\ Phys.\ Commun.\  {\bf 178}, 852 (2008)
  [arXiv:0710.3820 [hep-ph]].
  
\bibitem{deFavereau:2013fsa} 
  J.~de Favereau {\it et al.}  [DELPHES 3 Collaboration],
  JHEP {\bf 1402}, 057 (2014)
  [arXiv:1307.6346 [hep-ex]].  

\bibitem{Harlander:2012pb} 
  R.~V.~Harlander, S.~Liebler and H.~Mantler,
  Computer Physics Communications {\bf 184}, 1605 (2013)
  [arXiv:1212.3249 [hep-ph]].  
  
\bibitem{Campbell:2005zv} 
  J.~M.~Campbell, R.~K.~Ellis, F.~Maltoni and S.~Willenbrock,
  Phys.\ Rev.\ D {\bf 73}, 054007 (2006)
  [Erratum-ibid.\ D {\bf 77}, 019903 (2008)]
  [hep-ph/0510362].  

\bibitem{Mangano:1991jk} 
  M.~L.~Mangano, P.~Nason and G.~Ridolfi,
  Nucl.\ Phys.\ B {\bf 373}, 295 (1992);  
  G.~Bevilacqua, M.~Czakon, A.~van Hameren, C.~G.~Papadopoulos and M.~Worek,
  JHEP {\bf 1102}, 083 (2011)
  [arXiv:1012.4230 [hep-ph]].

\bibitem{Maltoni:2013sma} 
  F.~Maltoni, K.~Mawatari and M.~Zaro,
  Eur.\ Phys.\ J.\  {\bf 74}, 2710 (2014)
  [arXiv:1311.1829 [hep-ph]].  

\bibitem{Ohnemus:1994ff} 
  J.~Ohnemus,
  Phys.\ Rev.\ D {\bf 50}, 1931 (1994)
  [hep-ph/9403331].  
  
\bibitem{CMS:2013vea} 
  CMS Collaboration [CMS Collaboration],
  CMS-PAS-BTV-13-001.  
\bibitem{Lazopoulos:2008de}
  A.~Lazopoulos, T.~McElmurry, K.~Melnikov and F.~Petriello,
  Phys.\ Lett.\ B {\bf 666} (2008) 62
  [arXiv:0804.2220 [hep-ph]].
\end{thebibliography}
\end{document}